\documentclass[twoside,slac_one]{revtex4}
\usepackage{graphicx}
\usepackage{fancyhdr}
\usepackage{amsmath} 
\usepackage{bm}
\usepackage{amsxtra}
\usepackage{amssymb}
\usepackage{amsthm}
\usepackage{latexsym}
\usepackage{lscape}

\begin{document}

\title{MINER$\nu$A Detector: Description and Performance}

\author{B.Osmanov (on behalf of MINER$\nu$A collaboration)}
\affiliation{Department of Physics, University of Florida, Gainesville, FL, USA}

\begin{abstract}
The MINER$\nu$A experiment is aimed at precisely measuring the cross-sections for various neutrino interaction channels. It is located at Fermilab in the underground cavern in front of MINOS near detector. MINER$\nu$A is a fine-grained scintillator with  electromagnetic and hadronic calorimetry regions. There are various nuclear targets located inside and in front of the detector for studying nuclear medium effects in neutrino-induced interactions. The installation was completed in March 2010 and since then the detector has been collecting data. In this paper, the method for determining the neutrino flux is described in detail with the associated uncertainties as well as the techniques for their reduction. The general structure of the detector is given with the emphasis on the nuclear targets region. Preliminary results related to nuclear effects studies are presented followed by their discussion and future plans.

\end{abstract}

\maketitle

\thispagestyle{fancy}

\section{Introduction}
\begin{figure}[ht]
\centering
\includegraphics[width=80mm]{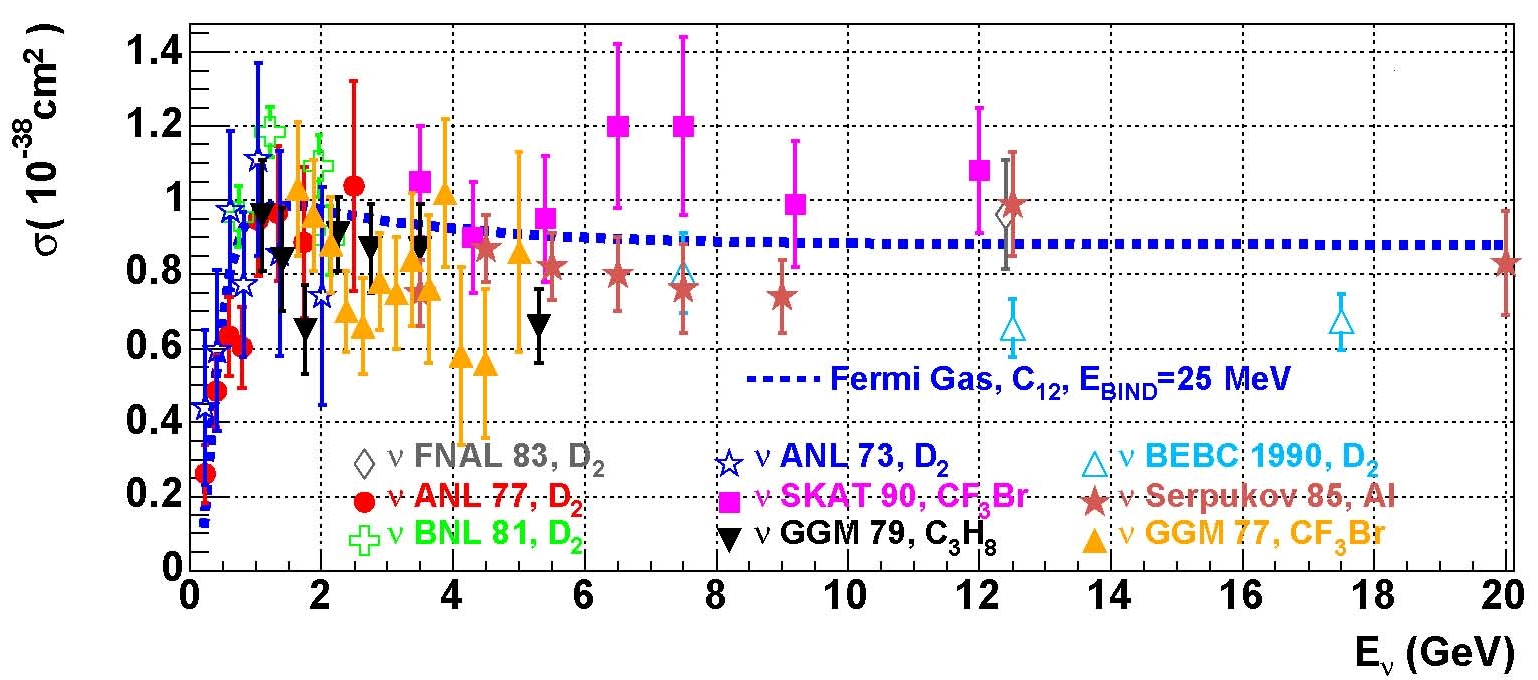}
\caption{Current measurements of total neutrino interaction cross-section compared to the prediction of Fermi gas model.}\label{figure1}
\end{figure}

The MINER$\nu$A collaboration emerged as a result of a joint effort between the high-energy physics and medium energy nuclear physics communities. The main goal of the experiment is to measure neutrino-nucleus interaction cross-section with a high degree of precision. This is an important issue for both present and future neutrino oscillation experiments. The current available cross-section measurements have large uncertainties due to low statistics and poor knowledge of the incoming neutrino flux as can be seen on ~Figure~\ref{figure1}.

\begin{figure}[ht]
\centering
\includegraphics[width=80mm]{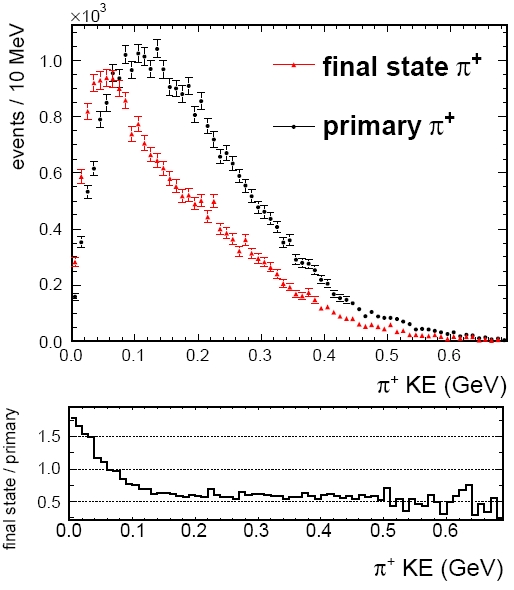}
\caption{Kinetic energy spectrum of final state and primary $\pi$$^{+}$  from $\nu$$_{\mu}$ + Fe56  interactions at 1 GeV ~\cite{genie}.}\label{figure2}
\end{figure}

MINER$\nu$A will collect large event samples for various interaction channels and will measure the cross-sections with negligible statistical errors and with the well-controlled beam systematic errors.
\\ \indent Another important goal of the experiment is to study nuclear effects in neutrino interactions. These effects such as final state interactions (as seen on ~Figure~\ref{figure2}) can modify the distribution of outgoing particles which will lead to the incorrect estimation of the incoming neutrino energy.

Various nuclear targets incorporated into the body of the detector as well as placed in front will allow to understand processes which are absent in neutrino scattering on free nucleons.

\section{Neutrino flux}
The NuMI beamline at Fermilab provides high-intensity neutrino beam with $\sim$35E12 protons per spill and $\sim$350 kW beam power ~\cite{numi} . The layout of the facility is shown on  ~Figure~\ref{figure2}. Main Injector produces 120 GeV proton beam that is incident on graphite target. The resulting  secondary hadrons are focused by the set of two magnetic horns. Following the horns, there is a decay pipe where the hadrons produce the neutrinos that enter the detector hall. Proton, hadron and muon monitors are installed throughout the beamline to have better control over the beam.

\begin{figure}[ht]
\centering
\includegraphics[width=135mm]{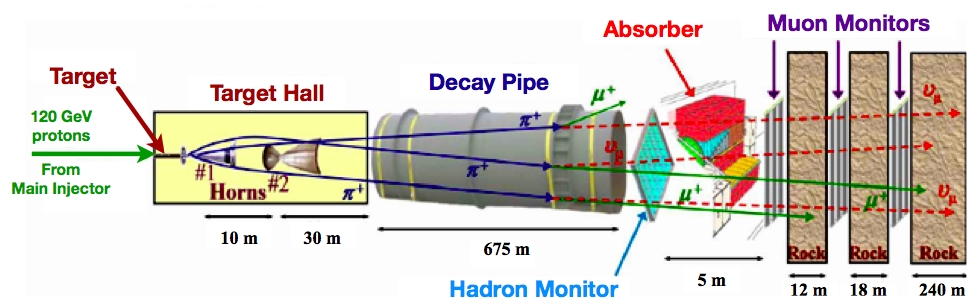}
\caption{NuMI beam facility layout.} \label{figure3}
\end{figure}

\begin{figure}[ht]
\centering
\includegraphics[width=135mm]{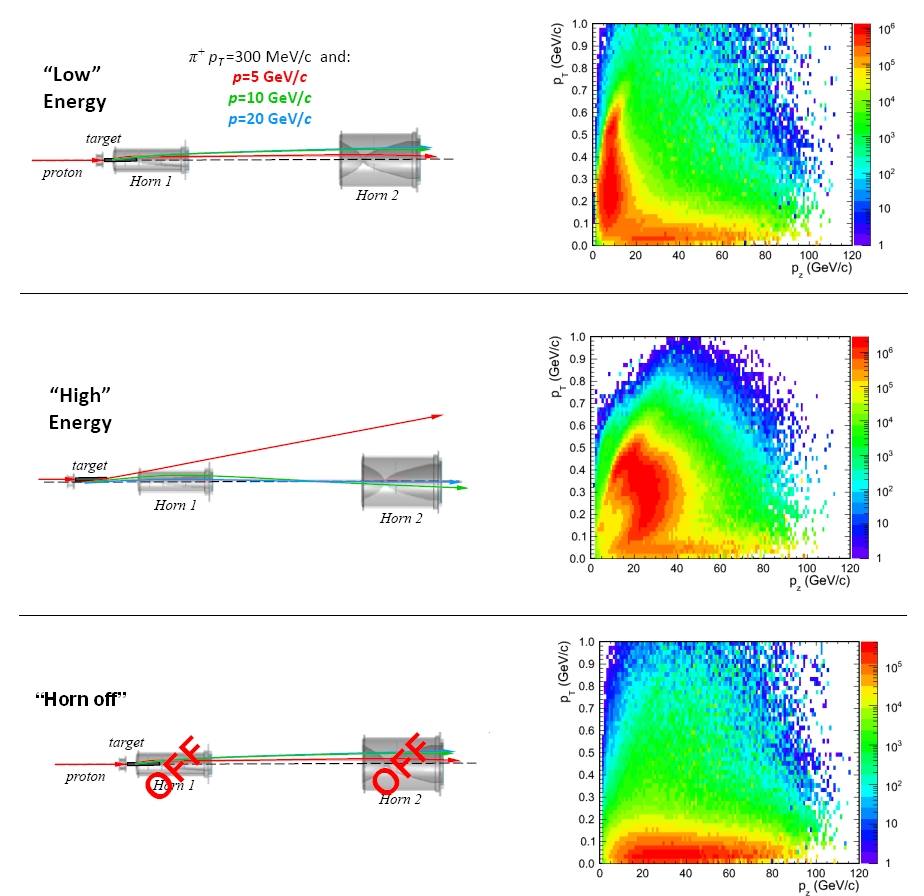}
\caption{Various energy modes of NuMI beamline.} \label{figure4}
\end{figure}

By changing the horn polarity one can switch between neutrino and anti-neutrino modes. Beam energy can be tuned by moving the target with respect to the first horn. As a result, pions and kaons of different energies reach the focusing region. ~Figure~\ref{figure4} illustrates this technique where different target-horn geometries are shown on the left with the resulting p$_{z}$/p$_{t}$ distributions of the produced hadrons on the right.

\begin{figure}[ht]
\centering
\includegraphics[width=80mm]{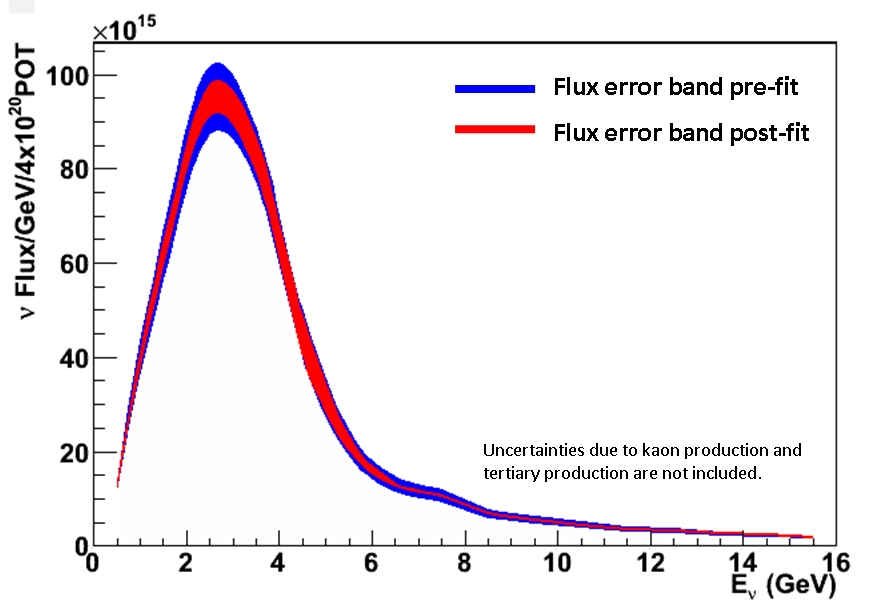}
\caption{Resulting reduction of errors from tuning hadron production yields.}  \label{figure5}
\end{figure}

It is known that the largest source of uncertainty in the predicted neutrino flux at the detector comes from the hadron production at the target. Our studies show that there is about 10\% spread in the peak of neutrino spectrum when the hadron production models are varied while keeping the geometry the same. MINER$\nu$A plans to use several techniques to reduce the flux-associated errors. First approach is to use the latest, more accurate external hadron production data. However, care must be taken since most of the available experimental data is for thin targets while NuMI target is two interaction lengths. In addition, one needs to take into account downstream interactions of the produced hadrons. We will also use the data from three muon monitors (shown on  ~Figure~\ref{figure4}) that are located at different distances from the target and thus sample different regions of neutrino energy spectrum. Finally, to better constrain hadron production at the target, a series of beam runs were taken with varying target position and horn current. Then, hadron production yields in MC are tuned to match the real data. The result is a set of weights in each of p$_{z}$-p$_{t}$ bins that is applied to MC pion yields. The outcome of this technique is shown on  ~Figure~\ref{figure5} where pre-fit errors are shown in blue and post-fit are in red.

\section{Nuclear targets and event estimates}
\begin{figure}[ht]
\centering
\includegraphics[width=135mm]{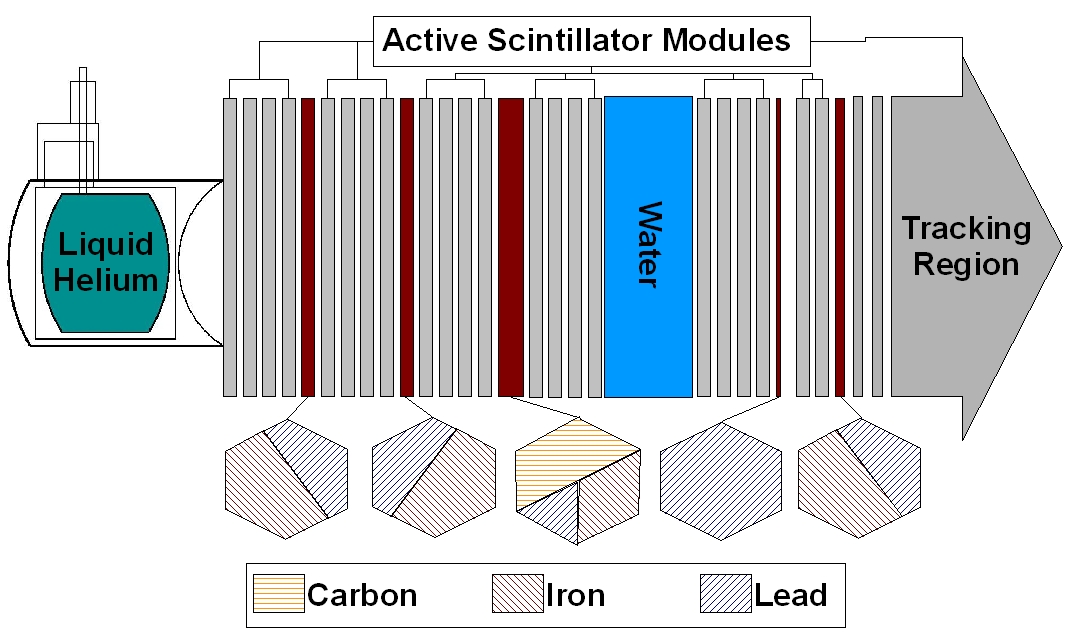}
\caption{Location of nuclear targets in MINER$\nu$A detector.} \label{figure6}
\end{figure}

MINER$\nu$A is a fine-grained scintillator-strip detector with electromagnetic and hadronic calorimetry regions. There are approximately 30,000 scintillator channels grouped into inner and outer regions. For details on general detector structure, see ~\cite{mcgowan}. 
\\ \indent For studying nuclear effects in neutrino interactions, various targets are incorporated into the body of the detector with interspersed layers of active scintillator. There is a liquid helium target located in front of the detector. The location of nuclear targets in shown on ~Figure~\ref{figure6}. All the targets are installed except for water target which is planned for installation later this year. Also, the collaboration has prepared a proposal for introducing additional nuclear target: deuterium for high-precision nuclear-to-D ratio measurements and comparison with the available charged lepton data as well as various theoretical models - that will replace the He.

The complete list of the targets with fiducial volumes and event estimates (with Genie 2.6.2) is given in ~Table~\ref{table1}.

\begin{table}[h]
\begin{center}
\caption{MINER$\nu$A nuclear targets with event estimates.}
\begin{tabular}{|l|c|c|}
\hline \textbf{Target} & \textbf{Fiducial mass} & \textbf{Numu CC events for 1.2E20 POT}\\
& \textbf{(tons)} & \textbf{(thousands)}\\
\hline Plastic & 6.43 & 409 \\
\hline Helium & 0.25 & 16.8 \\
\hline Carbon & 0.17 & 10.8 \\
\hline Water & 0.39 & 24.4 \\
\hline Iron & 0.97 & 64.5 \\
\hline Lead & 0.98 & 68.4 \\
\hline
\end{tabular}
\label{table1}
\end{center}
\end{table}

\section{Analysis chain}

For the initial analysis, the most downstream target (Fe/Pb) was chosen due to the ease of event reconstruction (no passive material in track's path). Charged current inclusive $\nu$$_{\mu}$ events were selected with one muon track matched to the  MINOS near-detector that we use to measure the muon momentum and charge. Low-energy neutrino mode (see Fig.4) was used with 0.9E20 and 11.2E20 protons-on-target for data and MC, respectively. The fiducial volume was a hexagon with 85 cm apothem with z position of muon vertex (reconstructed origin of muon track) in the nuclear target or the first module downstream. There is also a requirement of no muon-like activity upstream of the target. The resulting muon tracks are extrapolated backward into the target region and based on the projected location the event is classified as the one from iron or lead. At this analysis stage, the events were not background subtracted (the background that comes from the inability to resolve whether the event originated in the nuclear target or the first scintillator module downstream) so the samples are named “iron-enriched” and “lead-enriched”.

\section{Preliminary results}

\begin{figure}[ht]
\centering
\includegraphics[width=135mm]{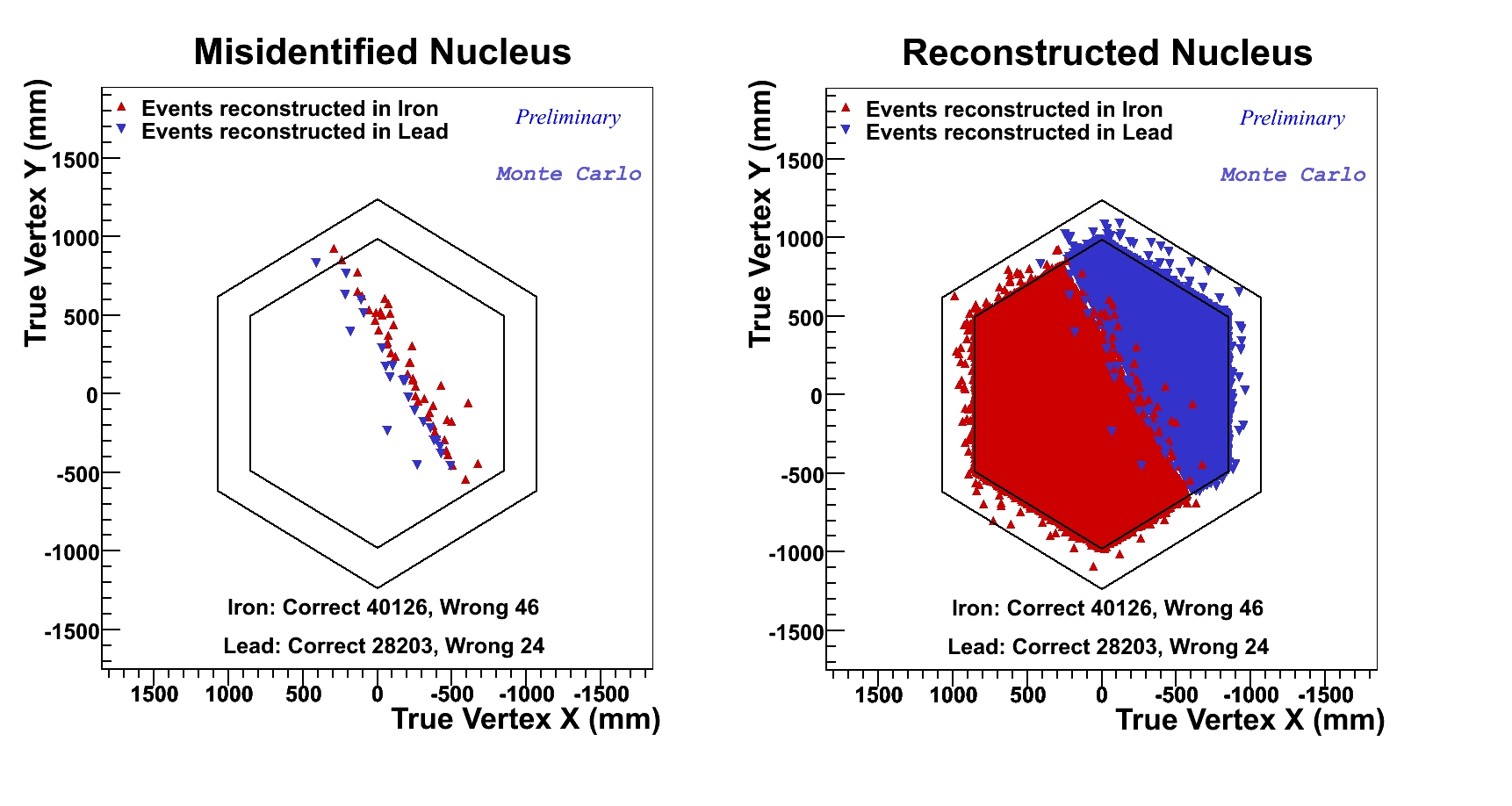}
\caption{Iron/lead event misidentification due to incorrect vertex reconstruction.} \label{figure7}
\end{figure}

Figure~\ref{figure7} shows the event misidentification due to incorrect reconstruction of the vertex where red triangles are true events in lead that were reconstructed in iron and vice versa for lead with blue triangles. As can be seen from the figure, vertex reconstruction works properly with only a few events being incorrectly vertexed.

\begin{figure}[ht]
\centering
\includegraphics[width=135mm]{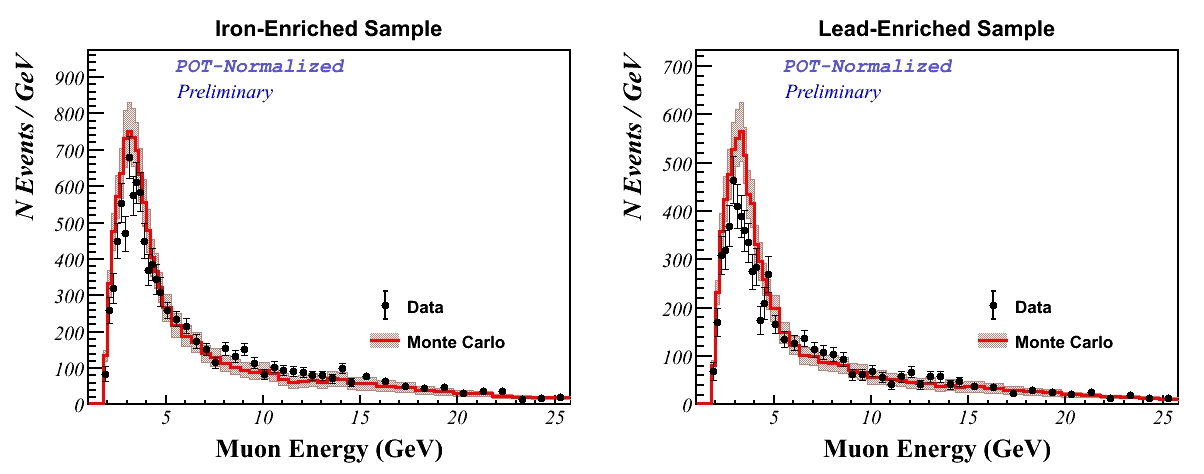}
\caption{Muon energy distributions for nuclear target.} \label{figure8}
\end{figure}

Data-MC comparison plots of muon energy for selected events are shown on ~Figure~\ref{figure8} for “iron-enriched” and “lead-enriched” samples.

\begin{figure}[ht]
\centering
\includegraphics[width=80mm]{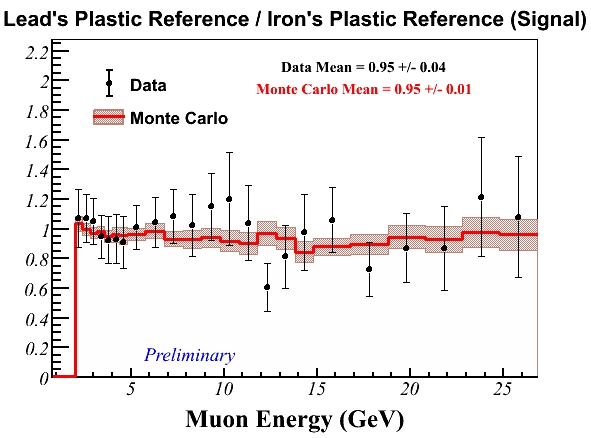}
\caption{Ratio of lead's to iron's plastic reference targets.}  \label{figure9}
\end{figure}

One of the scientific goals of nuclear target program of MINER$\nu$A experiment is to compare various reconstructed quantities (track multiplicities, etc) between nuclear targets and the active scintillator volume of the detector. For that purpose, a region with 4 scintillator modules (plastic reference target) was selected downstream of all the targets with the imaginary divide as in the last Fe/Pb target. This analysis is in progress and as a cross-check, the ratio of lead to iron plastic reference targets is shown on  ~Figure~\ref{figure9}. The reason this ratio is less than one is due to the fact that these two reference volumes occupy different regions in X-Y plane and thus have different acceptance values for muons propagating into MINOS detector.

The ratios among actual nuclear targets will be available in the nearest future. Final analysis will be done with approximately four times more protons-on-target and four times more fiducial mass which will provide a highly-precise measurement.

\begin{acknowledgments}
The support for this work was provided by the Fermi National Accelerator Laboratory operated by the Fermi Research Alliance, LLC, under contract No. DE-AC02-07CH11359 with the United States Department of Energy. Construction support also was granted by the United States National Science Foundation under NSF Award PHY-0619727 and by the University of Rochester.
\end{acknowledgments}

\end{document}